\documentclass[10pt,notitlepage]{iopart}
\usepackage{a4,epsfig}
\usepackage{url}
\usepackage[dvipsnames]{xcolor}
\usepackage[hidelinks]{hyperref}
\usepackage{lineno}
\usepackage{multirow}
\usepackage{xcolor}
\usepackage{cite}
\usepackage{iopams}
\usepackage[toc,page,title]{appendix}
\usepackage[normalem]{ulem}
\usepackage{graphicx}
\usepackage{subfig}
\begin{document}

\title{Relative calibration of the LIGO and Virgo detectors using astrophysical events from their third observing run}

\author{C.~All\'en\'e$^1$, N.~Andres$^1$, M.~Assiduo$^{2,3}$, F.~Aubin$^{2,3}$, D.~Buskulic$^1$, R.~Chierici$^4$, D.~Estevez$^5$\footnote{Corresponding author}, F.~Faedi$^{2,3}$, G.~M.~Guidi$^{2,3}$, V.~Juste$^5$, F.~Marion$^1$, B.~Mours$^5$, E.~Nitoglia$^4$, V.~Sordini$^4$, A.~Syx$^5$}

\address{$^1$Laboratoire d'Annecy de Physique des Particules (LAPP), Universit\'e Savoie Mont Blanc, CNRS/IN2P3, F-74941 Annecy, France}
\address{$^2$Universit\`a degli Studi di Urbino 'Carlo Bo,' I-61029 Urbino, Italy}
\address{$^3$INFN, Sezione di Firenze, I-50019 Sesto Fiorentino, Firenze, Italy}
\address{$^4$Institut de Physique des 2 Infinis de Lyon (IP2I) - UMR 5822, Universit\'e de Lyon, Universit\'e Claude Bernard, CNRS, F-69622 Villeurbanne, France}
\address{$^5$Universit\'e de Strasbourg, CNRS, IPHC UMR 7178, F-67000 Strasbourg, France}

\begin{abstract}
We explore a method to assess the relative scale of the strain measured in the different detectors of the gravitational-wave network, using binary black hole (BBH) events detected during the third observing run (O3).
The number of such signals is becoming sufficiently large to adopt a statistical approach based on the ratio of the signal-to-noise ratio (SNR) of the events between the detectors and the number of observed events in each detector.
We demonstrate the principle of the method on simulations of BBH signals and we present its application to published O3 events reported by the Multi-Band Template Analysis (MBTA) pipeline.
Constraints on the relative calibration of the gravitational-wave network for O3 are obtained at the level of $\sim3.5\%$ between the two LIGO detectors and at the level of $\sim10\%$ between the LIGO Livingston detector and the Virgo detector.
\end{abstract}

\ioptwocol
\clearpage 
\section{Introduction} \label{introduction}
With the prospects of the increasing number of gravitational-wave (GW) detections for the upcoming runs \cite{Prospects:2020},
efforts for precise and accurate calibration of GW detectors \cite{AdVLIGO:2015,AdVVirgo:2015} have been undertaken to set the calibration uncertainties at a level low enough not to be a limiting factor for scientific results \cite{TestGR:O3a,TestGR:O3b,AstroDist:O3a,AstroDist:O3b,Cosmo:O3}.
The level of uncertainty on the reconstructed GW strain amplitude achieved during O3 is $\sim 2\%$ for the LIGO detectors \cite{CalibLIGOO3:2020} and $5\%$ for the Virgo detector \cite{CalibVirgoO3:2022}.

The method used for the calibration of GW detectors during O3 relies on fiducial displacements of the test masses induced by auxiliary laser systems \cite{PCalLIGO:2021,PCalVirgo:2021}.
Another method based on the laser wavelength of the primary laser was also used during previous observing runs \cite{CalibFSMVirgo,CalibFSMLIGO} and can still be used for consistency checks of calibration.
Recently, an alternative technique based on test mass displacements using variations of the gravitational field induced by rotating masses, the so-called Newtonian calibrator (NCal), has also been investigated for Virgo \cite{NCalVirgo:2018,NCalVirgo:2021} and LIGO \cite{NCalLIGO:2021} and is being improved for the next observing runs with an expected subpercent uncertainty level.

An alternative approach using astrophysical sources to calibrate GW detectors has been investigated in recent years with various methods.
It follows a long tradition of using large population of real signals to calibrate the response uniformity of a detector, like it is done in high-energy physics detectors for instance \cite{CalibLHC:2018}.
Here, the goal is to cross-calibrate the network of GW detectors.
It has been shown in \cite{TimingCal:2009} with simulated data that calibration errors can be estimated using timing information in a GW detector with compact-binary coalescences (CBC).
In \cite{AstroCal:2016}, the consistency of amplitude scaling in individual detector calibration is assessed with simulations of joint detections of GW signals with short gamma ray-bursts.
An extension of this work on both absolute and relative responses of GW detectors with an application to GW170817 \cite{GW170817:2017} has also been demonstrated in \cite{GW170817Cal:2019}, assuming general relativity is correct.
The authors constrained the amplitude calibration of individual detectors with a precision of $\pm20\%$ around $100~$Hz using GW170817 only and down to $\pm10\%$ adding electromagnetic constraints on the luminosity distance and the orbital inclination.
Other investigations have been conducted to incorporate detailed calibration models in the inference of source parameters using detected events from the GWTC-1 publication \cite{GWTC1:2019} and to test on simulated events if these methods can inform calibration parameters \cite{AstroPhysical1:2020,AstroPhysical2:2021}.
Finally, the prospect of a self-calibration using the null stream of a GW network has also been proposed in \cite{SelfCal:2020}.

In this paper, we investigate and apply to LIGO and Virgo O3 data a method to measure the relative calibration between the GW detectors using the population of observed binary black hole (BBH) events.
An accurate relative calibration of the network is useful for sky localization.
Furthermore, it allows to transport the absolute calibration from one detector to another one and possibly improves the overall calibration of the network.

We first describe the method in section~\ref{method}, then a demonstration on simulations is presented in section~\ref{simulations} and finally we show the results on O3 data in section~\ref{results}.
\section{Method} \label{method}
The method we present in this paper is general and can be applied to the outcome of various searches. However, the following work has been performed using both simulated and real O3 data filtered with the Multi-Band Template Analysis (MBTA) pipeline \cite{MBTA:2021}, which produces coincident events when a signal is seen for the same template in multiple detectors.

\subsection{Impact of a calibration scaling error}
The measurement of the detector strain (which may contain a GW signal) is performed using differential arm length variations of the interferometer, $\Delta L$, and is expressed as $d(f,t) = \Delta L(f,t)/L$, where $L$ is the arm length, $f$ the frequency of the variations and $t$ the time.
In practice, $\Delta L$ is measured through an error signal output $d_{err}$ and both quantities are related via the calibration response function $R(f,t)$ as $\Delta L(f,t) = R(f,t)d_{err}(f,t)$.

In general, $R(f,t)$ could differ from the true response function due to a calibration error which could be time and frequency dependent \cite{CalibLIGOO3:2020}. However, due to the currently limited statistics of astrophysical GW events, we investigate the time- and frequency-independent part of the calibration error, coming in the form of a scaling factor $C$.
The output signal of a detector can be written in the frequency domain as:
\begin{equation}
d(f) = C[n(f)+h(f)]
\end{equation}
with $n(f)$ and $h(f)$ the Fourier transforms of the noise time series and a GW signal.
The signal-to-noise ratio (SNR) in a detector, for a matched-filter based search, is defined as:
\begin{equation}
\rho \equiv \frac{\langle d|T \rangle}{\sqrt{\langle T|T \rangle}}
\label{eq:snr}
\end{equation}
where $T$ is a GW waveform template from the search template bank, and the noise-weighted inner product is:
\begin{equation}
\langle a|b \rangle \equiv 4 \Re\Big[\int_{f_{\mathrm{min}}}^{f_{\mathrm{max}}} \frac{a(f)b(f)^*}{S_{n}(f)}df\Big]
\end{equation}
\begin{figure*} [h]
\centerline{\includegraphics[trim={0cm 1cm 0cm 1cm},clip,width=1.1\textwidth]{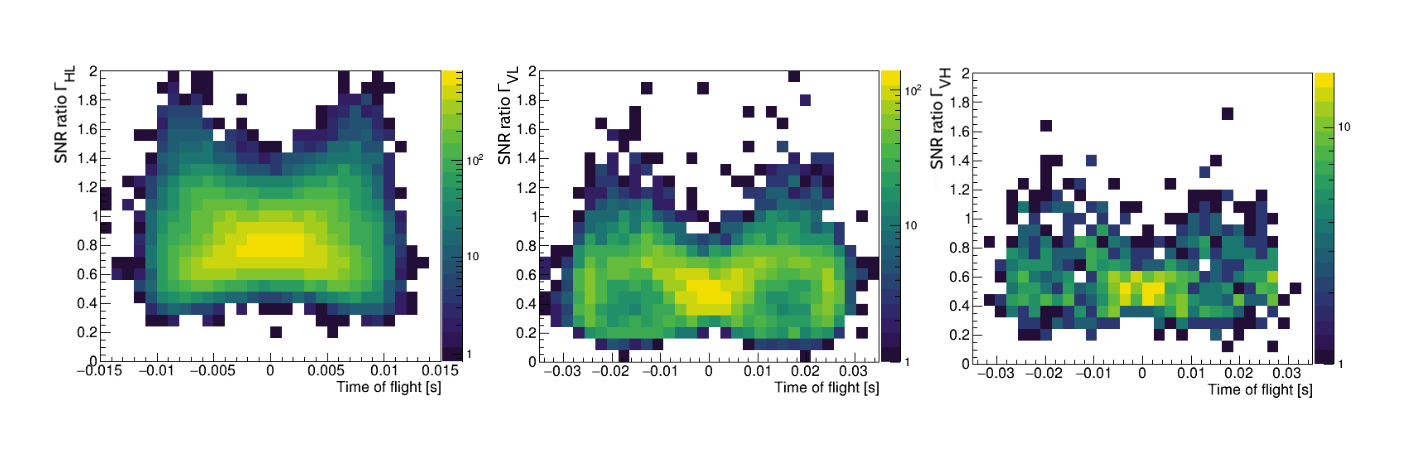}}
\caption{Joint distributions of the SNR ratio and the time of flight between two detectors for the BBH injections. The three types of coincidences using LIGO Hanford and LIGO Livingston (HL), Virgo and LIGO Livingston (VL) and Virgo and LIGO Hanford (VH) are shown from left to right. The total number of recovered injections per coincidence type is 45650 for HL, 9145 for VL and 994 for VH. \label{fig:distrib_injections}}
\end{figure*}
with $S_{n}(f)$ the one-sided power spectral density of the detector noise which is also affected by calibration errors as:
\begin{equation}
S_n(f) = C^2 ~S_{n,\mathrm{true}}(f)
\end{equation}
where $S_{n,\mathrm{true}}$ is the true power spectral density without any calibration errors.
Hence, by construction, the SNR of candidate events is independent of $C$.

If we now consider simulated GW signals $h_{\mathrm{sim}}$, that we call \textit{injections}, which are directly added, without scaling factor, to the reconstructed detector strain, the data can be expressed as:
\begin{equation}
d_{\mathrm{sim}}(f) = C\cdot n(f)+h_{\mathrm{sim}}(f)
\end{equation}
From equation \ref{eq:snr} applied to the simulated data $d_{\mathrm{sim}}$, when $\langle n|T \rangle \ll \langle h_{\mathrm{sim}}|T \rangle$, the SNR of unambiguously recovered injections is proportional to $C^{-1}$.

In what follows, we first perform injections by adding $h_{\mathrm{sim}}(f)$ to a perfectly-calibrated, imaginary detector network with true noise $C \cdot n(f)$ (i.e. the same as the apparent noise of the imperfectly-calibrated, real detector network), and use them as a reference for how signals should appear in perfectly-calibrated detectors. We then try to find the relative scaling factors that need to be applied to $h_{\mathrm{sim}}(f)$ so that injections behave in the same way as real events in the real detector network, as a way to extract relative calibration errors in the latter.

We define the SNR ratio for each pair of detectors as $\Gamma_{\mathrm{HL}} = \rho_{\mathrm{H}}/\rho_{\mathrm{L}}$, $\Gamma_{\mathrm{VL}} =\rho_{\mathrm{V}}/\rho_{\mathrm{L}}$ and $\Gamma_{\mathrm{VH}} = \rho_{\mathrm{V}}/\rho_{\mathrm{H}}$ with LIGO Livingston (L), LIGO Hanford (H) and Virgo~(V).
We also note the relative calibration factors as $C_{\mathrm{HL}}$, $C_{\mathrm{VL}}$ and $C_{\mathrm{VH}}$, which are inversely proportional to the SNR ratio for the injections.
Using the distribution of SNR ratio measured on simulated and real data for each pair of detectors is thus a way to estimate the relative calibration factor between them.

\subsection{Using simulated GW signals}
The method we investigate compares the joint distribution of GW events detected in the LIGO and Virgo data in terms of SNR ratio $\Gamma_{\mathrm{AB}}$ and time of flight $\Delta t_{\mathrm{AB}}$ between two detectors $\mathrm{A}$ and $\mathrm{B}$ ($\mathrm{AB}\in\{\mathrm{HL}, \mathrm{VL}, \mathrm{VH}\}$) to a fiducial distribution of simulated GW signals free of calibration errors.
The time of flight is used to correlate the detections with the sky location and therefore the expected SNR ratio.
For this study, we use the same reweighted BBH injections as those used for the computation of the probability of astrophysical origin of candidate events detected with the MBTA pipeline for O3 \cite{MBTA:2021,pastroMBTA:2022}, included in the GWTC-2.1 and GWTC-3 catalogs \cite{GWTC2.1:2021, GWTC3:2021}. The assumed population follows a uniform distribution in comoving volume without any redshift evolution, the \textsc{Power Law + Peak} distribution for the masses as inferred in \cite{AstroDist:O3a} and an isotropic distribution for the spins orientation.
We emphasize that the population assumptions should not impact significantly the distribution of SNR ratio and time of flight since the ratio depends mostly on the relative sensitivities of the detectors and the time of flight distribution depends on the isotropy of the injections.
We select injections recovered with the MBTA pipeline using a threshold on the probability of astrophysical origin of $p_{\mathrm{astro}}>0.5$ and we also apply a cut on the SNR in the most sensitive detector ($\mathrm{SNR}\geq7$) to avoid counting any noise event randomly associated with an injection.

Although in principle it is also possible to use the time of flight to assess the relative phase calibration errors, the number of events detected during O3 is not sufficient to obtain significant constraints (see \ref{appendix} for more details).

Since the goal of the method is to estimate the relative calibration factors for each pair of detectors, we split the recovered injections into the three coincidence types.
We consider that events occur as independent Poisson processes for each type of coincidence.
As this assumption is only true for double coincidences, we count the triple detector coincidences ``HLV" as part of the HL and VL distributions only, in order to keep the independence between the coincidence types.
Figure \ref{fig:distrib_injections} shows the joint distribution in $\Gamma_{\mathrm{AB}}$ and $\Delta t_{\mathrm{AB}}$ of recovered BBH injections for the three types of coincidences using unscaled injections ($C_{\mathrm{HL}} = C_{\mathrm{VL}} = 1$).
As expected, the average value of $\Gamma_{\mathrm{HL}}$ is closer to $1$ than for the other types of coincidences as the sensitivities of the LIGO detectors are similar.
Moreover, the joint distributions are symmetric with respect to a zero time of flight ($\Delta t_{\mathrm{AB}} = 0$).
This feature allows us to use the absolute value of the time of flight of the recovered injections and thus increase the statistics in each bin of the joint distributions.

\subsection{Relative calibration using the SNR ratio}

We aim at inferring the relative calibration factors using a maximum likelihood estimation (MLE).
To do so, we consider a first Poisson likelihood expressed as:

\begin{equation}
\resizebox{0.49\textwidth}{!}{$\mathcal{L}_1(\vec{k}|C_{\mathrm{HL}},C_{\mathrm{VL}}) = \displaystyle\prod_{\mathrm{AB}} \displaystyle\prod_{i,j} e^{-\lambda_{i,j}(C_{\mathrm{AB}})}\frac{\lambda_{i,j}(C_{\mathrm{AB}})^{k^{\mathrm{AB}}_{i,j}}}{k^{\mathrm{AB}}_{i,j}!}$}
\end{equation}
with $\vec{k}=\{k^{\mathrm{AB}}_{i,j}\}$ the list of numbers of events for coincidence types $\mathrm{AB}$ detected in bins $(i,j)$ of the $(\Gamma_{\mathrm{AB}},\Delta t_{\mathrm{AB}})$ space over a given period of time and $\lambda_{i,j}(C_{\mathrm{AB}})$ the expected number of events in bin $(i,j)$ given the relative calibration factor $C_{\mathrm{AB}}$.
We stress that the relative calibration factors $C_{\mathrm{HL}}$ and $C_{\mathrm{VL}}$ are treated as independent but $C_{\mathrm{VH}}$ is a combination of the other two factors $C_{\mathrm{VH}} =~C_{\mathrm{VL}}/C_{\mathrm{HL}}$. 

The expected numbers of events $\lambda_{i,j}(C_{\mathrm{AB}})$ are directly estimated from the joint distributions of injections by multiplying $\Gamma_{\mathrm{AB}}$ with different values of $C_{\mathrm{AB}}$, which gives a new distribution for each of those values.
However, some bins are empty or are subject to large statistical fluctuations because the number of injections is limited as shown in figure \ref{fig:distrib_injections}.
In practice, the $\Delta t_{\mathrm{AB}}$ histograms are first smoothed with a kernel ($15$ histograms between $[0,15~$ms] for HL types and $15$ histograms between $[0,35$~ms] for VL types). 
Then, we consider bins of width $0.01$ in $\Gamma_{\mathrm{AB}}$ to compute the $\lambda_{i,j}(C_{\mathrm{AB}})$.
The expected number of events for O3 in bin $(i,j)$ for a given $C_{\mathrm{AB}}$ is calculated as:
\begin{equation}
\lambda_{i,j}(C_{\mathrm{AB}}) = \frac{N^{\mathrm{rec}}_{i,j}(C_{\mathrm{AB}})}{\sum_{p,q}N^{\mathrm{rec}}_{p,q}(C_{\mathrm{AB}})} K_{\mathrm{AB}}
\end{equation}
with $N^{\mathrm{rec}}_{i,j}(C_{\mathrm{AB}})$ the number of recovered injections in bin $(i,j)$ given $C_{\mathrm{AB}}$ and $K_{\mathrm{AB}}$ the number of events detected during O3 for the coincidence type $\mathrm{AB}$.

\subsection{Relative calibration using the number of events}
We also consider a second Poisson likelihood carrying the information on the number of events per coincidence type among the entire set of events such that:
\begin{equation}
\resizebox{0.49\textwidth}{!}{$\mathcal{L}_2(\vec{K}|C_{\mathrm{HL}},C_{\mathrm{VL}}) = \displaystyle\prod_{\mathrm{AB}} e^{-\Lambda_{\mathrm{AB}}(C_{\mathrm{HL}}, C_{\mathrm{VL}})}\frac{\Lambda_{\mathrm{AB}}(C_{\mathrm{HL}},C_{\mathrm{VL}})^{K_{\mathrm{AB}}}}{K_{\mathrm{AB}}!}$}
\end{equation}
with $\vec{K}=\{K_{\mathrm{AB}}\}$ the list of numbers of events for coincidence types $\mathrm{AB}$ and $\Lambda_{\mathrm{AB}}(C_{\mathrm{HL}},C_{\mathrm{VL}})$ the expected number of events of coincidence type $\mathrm{AB}$ given $C_{\mathrm{HL}}$ and $C_{\mathrm{VL}}$.
\begin{figure*} [h]
\centerline{\includegraphics[trim={0cm 0cm 0cm 0cm},clip,width=1.03\textwidth]{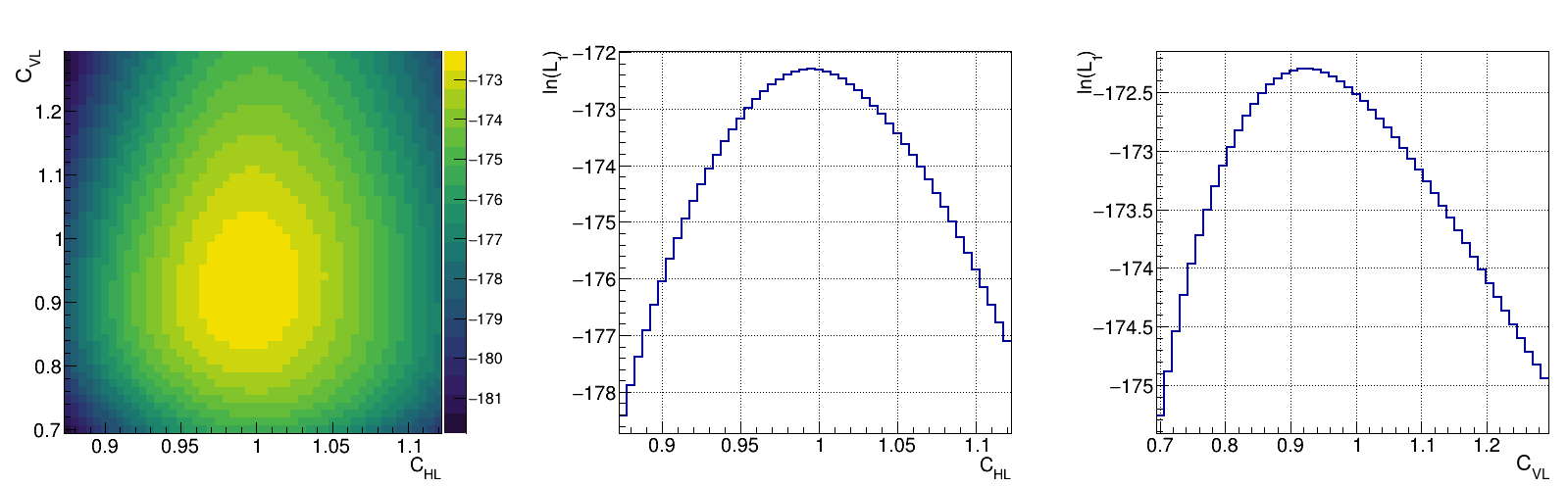}}
\caption{Log-likelihood $\ln(\mathcal{L}_1)$ of the relative calibration factors $C_{\mathrm{HL}}$ and $C_{\mathrm{VL}}$ for one specific simulation without any calibration errors ($C_{\mathrm{HL}}=C_{\mathrm{VL}}=1$). The profile log-likelihoods are also shown. \label{fig:loglikeli1_MC15}}
\end{figure*}
The expected numbers of events $\Lambda_{\mathrm{AB}}(C_{\mathrm{HL}}, C_{\mathrm{VL}})$ are directly estimated from the number of AB recovered injections $N^{\mathrm{rec}}_{\mathrm{AB}}(C_{\mathrm{HL}},C_{\mathrm{VL}})$ when applying different calibration factors on the SNR thresholds to select triggers.
Indeed, changing the SNR thresholds is equivalent to making the observed volume vary, which leads to different numbers of detections.
In practice, there is a minimal SNR threshold of 4.8 for each detector set by the MBTA search below which coincident triggers are not stored \cite{MBTA:2021}.
Hence, to explore $C_{\mathrm{HL}}$ and $C_{\mathrm{VL}}$ values below or above 1, the SNR threshold is increased either in one or the other two detectors.
The fraction of recovered injections of each coincidence type among the total number of recovered injections is then rescaled by the total number of events detected during O3 to get the expected number of events per coincidence type, such as:
\begin{equation}
\Lambda_{\mathrm{AB}}(C_{\mathrm{HL}},C_{\mathrm{VL}}) = \frac{N^{\mathrm{rec}}_{\mathrm{AB}}(C_{\mathrm{HL}},C_{\mathrm{VL}})}{\sum_{\mathrm{CD}}N^{\mathrm{rec}}_{\mathrm{CD}}(C_{\mathrm{HL}},C_{\mathrm{VL}})} \sum_{\mathrm{CD}} K_{\mathrm{CD}}
\end{equation}
where CD runs on the three types of coincidences.
Eventually, the likelihood we want to maximize to estimate the relative calibration factors is:
\begin{equation}
\resizebox{0.49\textwidth}{!}{$\mathcal{L}(\vec{k},\vec{K}|C_{\mathrm{HL}},C_{\mathrm{VL}}) = \mathcal{L}_1(\vec{k}|C_{\mathrm{HL}},C_{\mathrm{VL}}) \cdot \mathcal{L}_2(\vec{K}|C_{\mathrm{HL}},C_{\mathrm{VL}})$}
\end{equation}
\section{Simulations} \label{simulations}
Before applying our method to the O3 detections, we test it on simulations.
\begin{figure*}[h]
  \centering
  \subfloat[]{\includegraphics[trim={0cm 0cm 0cm 1cm},clip,width=0.79\textwidth]{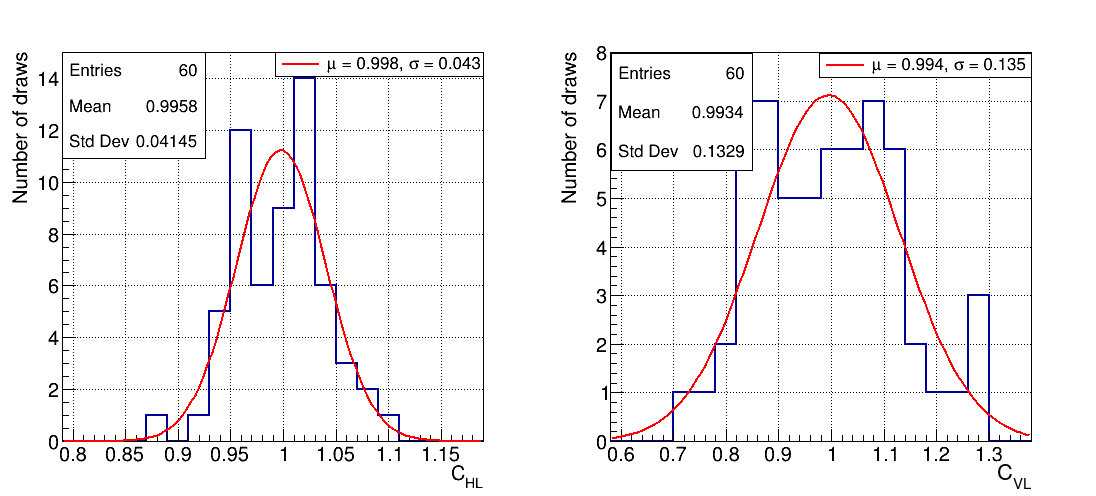}\label{fig:distrib_calfactors_nobias}}
  \vfill
  \subfloat[]{\includegraphics[trim={0cm 0cm 0cm 1cm},clip,width=0.79\textwidth]{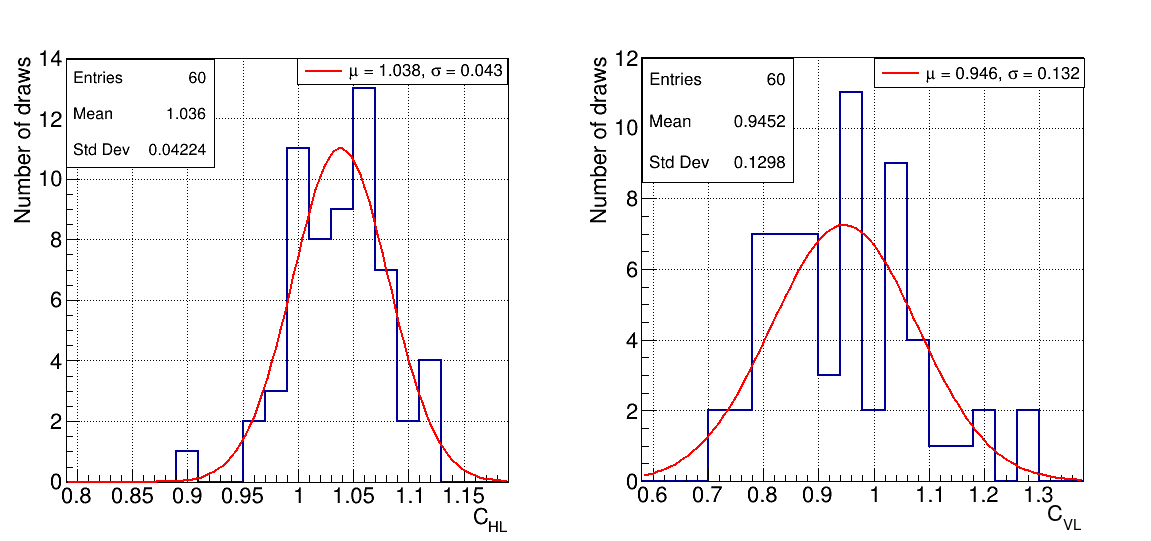}\label{fig:distrib_calfactors_bias}}
  \caption{Histograms of the relative calibration factors $C_{\mathrm{HL}}$ and $C_{\mathrm{VL}}$ for $60$ independent simulations of the O3 scenario (a) without any calibration errors ($C_{\mathrm{HL}}=1$ and $C_{\mathrm{VL}}=1$), (b) with calibration errors ($C_{\mathrm{HL}}=1.04$ and $C_{\mathrm{VL}}=0.95$) using $\mathcal{L}_1$. A gaussian fit is performed (red curve) on the histograms and the mean $\mu$ and standard deviation $\sigma$ are given.}
\end{figure*}
\begin{figure*}[h]
  \centering
  \subfloat[]{\includegraphics[trim={0cm 0cm 0cm 1cm},clip,width=0.4\textwidth]{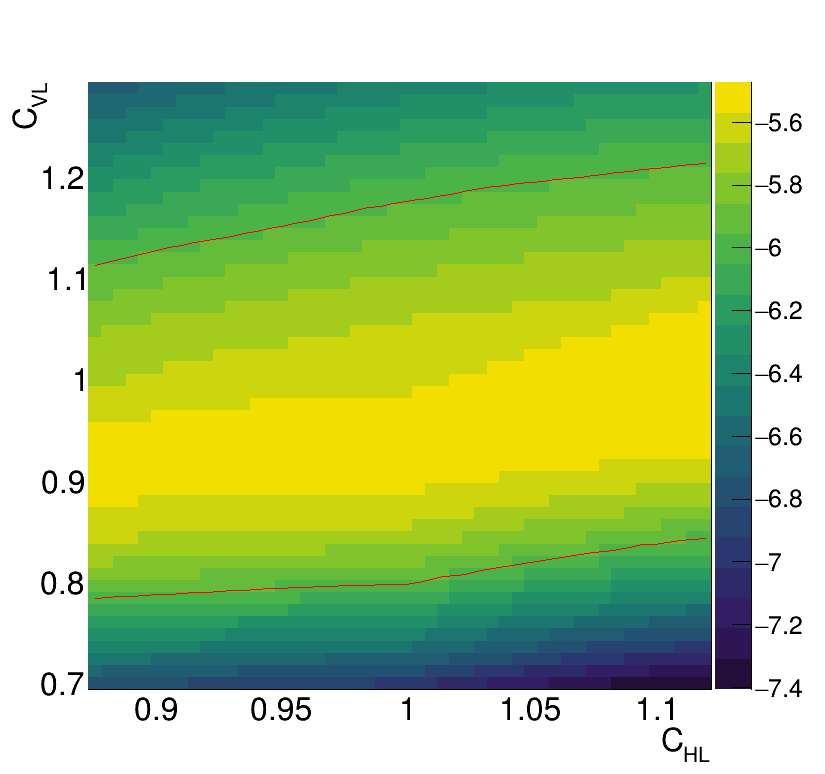}\label{fig:fit_CHL_CLV_MCcounts_method_nobias}}
  \hfill
  \subfloat[]{\includegraphics[trim={0cm 0cm 0cm 1cm},clip,width=0.4\textwidth]{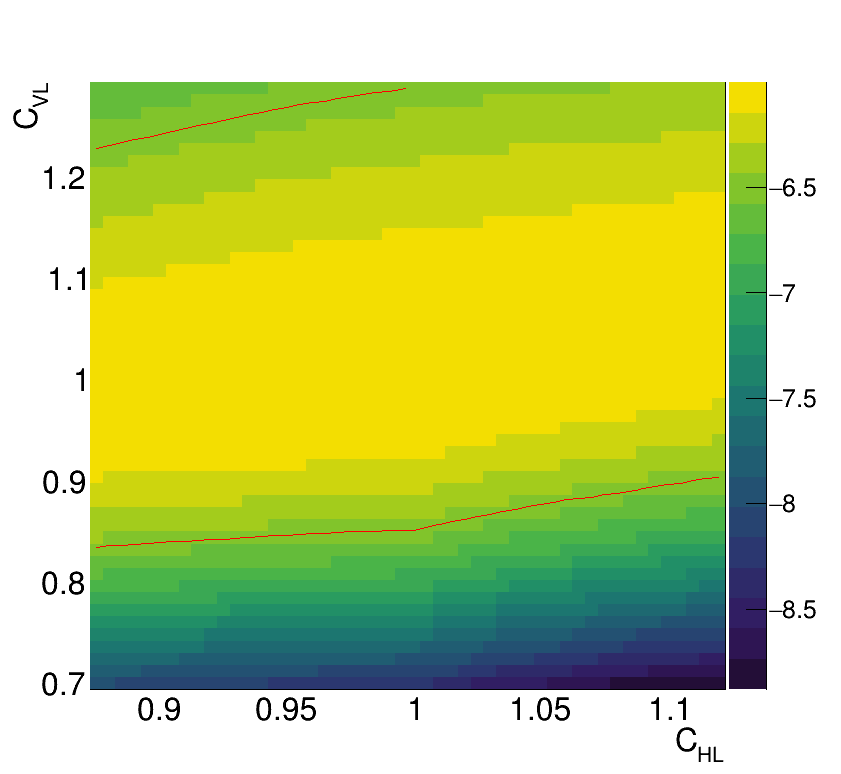}\label{fig:fit_CHL_CLV_MCcounts_method_bias}}
  \caption{(a) Log-likelihood $\ln(\mathcal{L}_2)$ of the relative calibration factors $C_{\mathrm{HL}}$ and $C_{\mathrm{VL}}$ for the same simulation as shown in figure~\ref{fig:loglikeli1_MC15}, computed using the number of selected MBTA BBH events with initial values set to $C_{\mathrm{HL}}=C_{\mathrm{VL}}=1$. The $1\sigma$ contour is drawn in red. (b) Log-likelihood $\ln(\mathcal{L}_2)$ for the relative calibration factors $C_{\mathrm{HL}}$ and $C_{\mathrm{VL}}$ averaged over $60$ simulations computed using the number of selected MBTA BBH events with initial values set to $C_{\mathrm{HL}}=C_{\mathrm{VL}}=1$. The $1\sigma$ contour is drawn in red.}
\end{figure*}
To do this, we first consider the likelihood $\mathcal{L}_1$ on the SNR ratio and the time of flight.
We draw $60$ independent scenarios in the BBH simulated signals without any calibration errors ($C_{\mathrm{HL}}=1$ and $C_{\mathrm{VL}}=~1$).
Each one of these scenarios contains the same numbers of coincident detections as the ones reported by MBTA in the O3 catalogs passing the selection criteria defined in section \ref{method} (i.e. 32 HL, 5 VL and 1 VH). The simulated coincident detections are sampled uniformly in time over O3.
Then, we perform a MLE for each draw, after removing the simulated detections from the set of BBH injections used to build the expected distribution of SNR ratio and time of flight.
The MLE is done over values of $C_{\mathrm{HL}}\in~[0.875,1.125]$ and $C_{\mathrm{VL}}\in~[0.70,1.30]$.

In figure \ref{fig:loglikeli1_MC15} we show an example of a typical log-likelihood $\ln(\mathcal{L}_1)$ for one of the simulations.
In this specific case, the relative calibration factors are estimated to $C_{\mathrm{HL}} = 0.995 \pm 0.035$ and $C_{\mathrm{VL}} =~0.928^{+0.120}_{-0.108}$, with the $1\sigma$ uncertainty on the relative calibration factors computed from the profile log-likelihoods as $\ln(\mathcal{L}_1)_{\mathrm{max}}-0.5$ \cite{LikelihoodUncertainty:2018}.

Repeating the MLE for the $60$ simulations, we show on figure \ref{fig:distrib_calfactors_nobias} the histograms of the relative calibration factors $C_{\mathrm{HL}}$ and $C_{\mathrm{VL}}$ corresponding to the maximum likelihood values computed for the $60$ scenarios.
The mean relative calibration factors found are $\langle  C_{\mathrm{HL}} \rangle = 0.996 \pm 0.005$ and $\langle  C_{\mathrm{VL}} \rangle =~0.993 \pm 0.017$ where the quoted uncertainties assume 60 independent draws.
The relative calibration factors are thus compatible with the injected value of $1$.
The standard deviation of the distributions are also comparable to the $1\sigma$ uncertainties estimated for one simulation.
Moreover, we performed a Gaussian fit of these histograms to compare them to normal distributions.

To show that the results are consistent on a more general case, we also performed a similar analysis on the same set of simulated events but with relative calibration factors set to $C_{\mathrm{HL}}=1.04$ and $C_{\mathrm{VL}}=0.95$ applied to their recovered SNR ratio.
In figure \ref{fig:distrib_calfactors_bias} we show similar distributions as in figure~\ref{fig:distrib_calfactors_nobias}.
This time, the mean relative calibration factors found are $\langle C_{\mathrm{HL}} \rangle = 1.036 \pm 0.005$ and $\langle C_{\mathrm{VL}} \rangle = 0.945 \pm 0.017$. 
These results are also compatible with the injected values of calibration factors, which confirms the validity of the method.

Then, we consider the likelihood $\mathcal{L}_2$ on the number of events.
We first use the same simulation as the one illustrated in figure~\ref{fig:loglikeli1_MC15} with initial values set to $C_{\mathrm{HL}} = C_{\mathrm{VL}} = 1$ and we show the results in figure~\ref{fig:fit_CHL_CLV_MCcounts_method_nobias}. We also perform $60$ independent simulations by considering the same number of events as for the real case, i.e. $38$ events. 
We draw them randomly from the BBH recovered injections without any constraint on the number of injections per coincidence type.
Then, we make the average of the 60 log-likelihoods for these calibration factors and we show the results in figure~\ref{fig:fit_CHL_CLV_MCcounts_method_bias}.
$\mathcal{L}_2$ is almost uninformative for $C_{\mathrm{HL}}$, which is due to the fact that the fraction of HL coincidences dominates the total number of events, and is therefore weakly sensitive to $C_{\mathrm{HL}}$.
However, the uncertainty on $C_{\mathrm{VL}}$ is of the same order of magnitude (around $\pm0.15$) as the one computed with $\mathcal{L}_1$, meaning that $\mathcal{L}_2$ could be useful to better constrain the $C_{\mathrm{VL}}$ value.

\begin{figure*} [h]
\centerline{\includegraphics[trim={0cm 0cm 0cm 0cm},clip,width=1.03\textwidth]{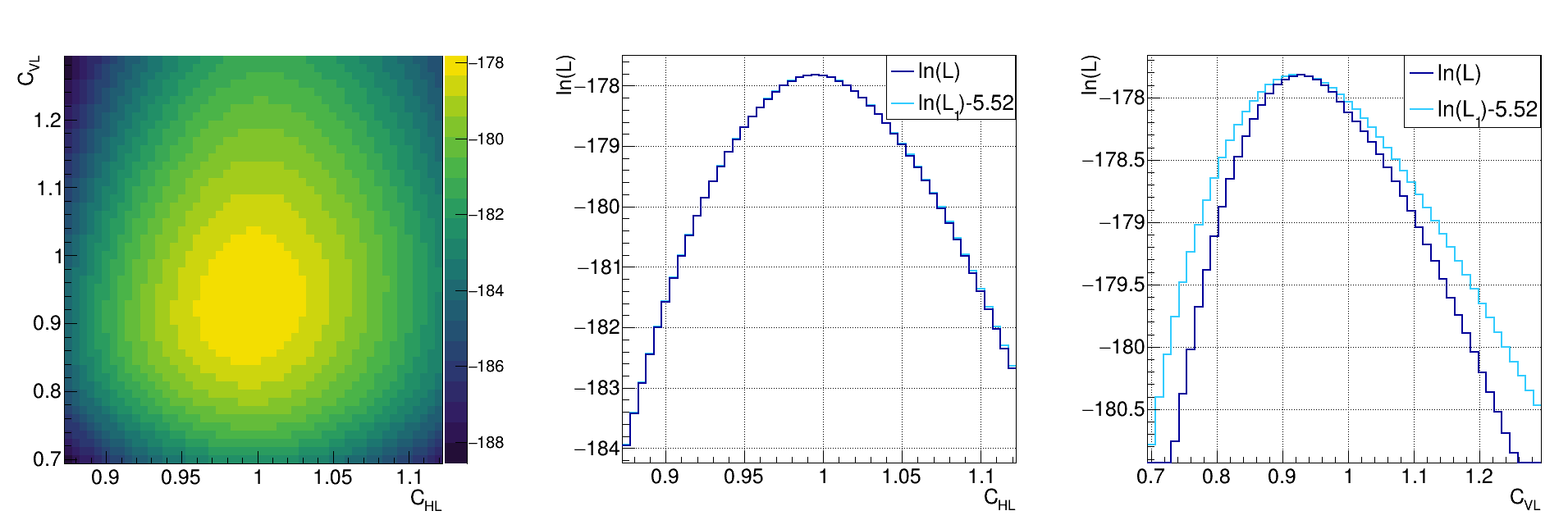}}
\caption{Log-likelihood $\ln(\mathcal{L})$ of the relative calibration factors $C_{\mathrm{HL}}$ and $C_{\mathrm{VL}}$ for the same simulation as in figure~\ref{fig:loglikeli1_MC15}. The profile log-likelihoods are shown with the addition of $\ln(\mathcal{L}_1)$ (cyan curves) with an arbitrary offset. An improvement on the uncertainty is visible for $C_{\mathrm{VL}}$. \label{fig:loglikeli_MC15}}
\end{figure*}

To illustrate the effect of $\mathcal{L}_2$ on the estimation of the relative calibration factors, we show $\ln(\mathcal{L})$ in figure~\ref{fig:loglikeli_MC15} for the same specific simulation used in figures~\ref{fig:loglikeli1_MC15} and \ref{fig:fit_CHL_CLV_MCcounts_method_nobias}.
As expected, $C_{\mathrm{HL}}$ is unchanged with respect to the values found with $\mathcal{L}_1$ only, but the value of $C_{\mathrm{VL}}$ has a smaller uncertainty $C_{\mathrm{VL}} = 0.928^{+0.108}_{-0.096}$.

In conclusion, the method we propose has been validated using simulations and we expect the global likelihood $\mathcal{L}$ to be mostly informed by the SNR ratio used in $\mathcal{L}_1$, with a slight improvement on the uncertainty of $C_{\mathrm{VL}}$ due to the number of events used in $\mathcal{L}_2$.
\section{Results on O3 data} \label{results}
\begin{figure} [h]
\centerline{\includegraphics[trim={0cm 0cm 1cm 1cm},clip,width=0.48\textwidth]{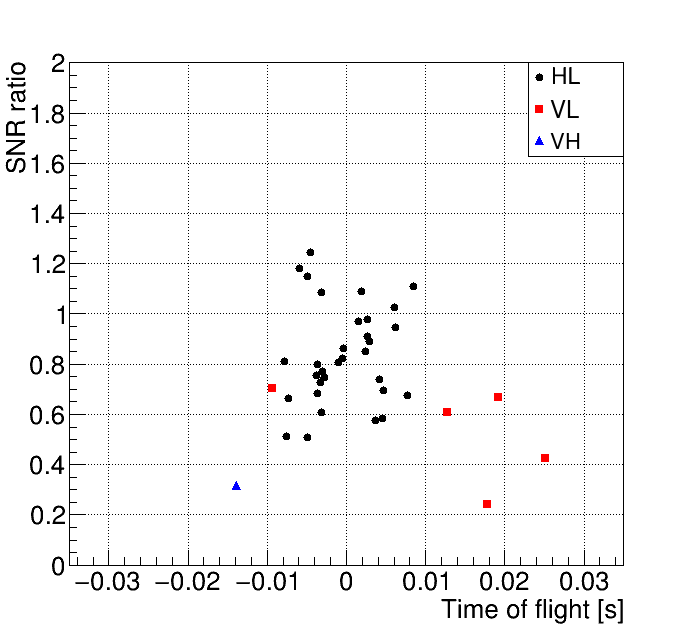}}
\caption{Joint distributions of SNR ratio and time of flight between two detectors for the selected BBH events detected with MBTA during O3. The three types of coincidences HL (black dots), VL (red squares) and VH (blue triangle) are shown.\label{fig:distrib_events}}
\end{figure}

We apply our method to the O3 MBTA events selected with $p_{\mathrm{astro}}>0.5$ and a SNR cut of $7$ on the SNR in the most sensitive detector.
The distribution of those events in $\Gamma_{\mathrm{AB}}$ and $\Delta t_{\mathrm{AB}}$ is shown in figure \ref{fig:distrib_events}.
\begin{figure*}[h]
\centerline{\includegraphics[trim={0cm 0cm 0cm 0cm},clip,width=1.03\textwidth]{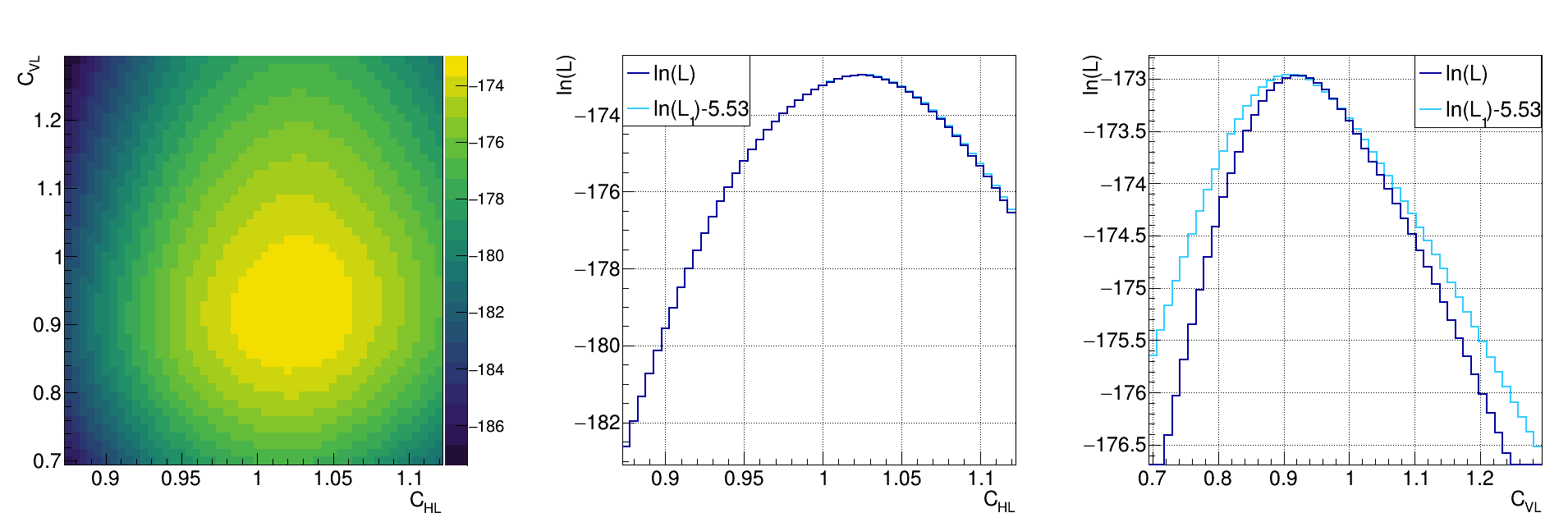}}
\caption{Log-likelihood $\ln(\mathcal{L})$ of the relative calibration factors $C_{\mathrm{HL}}$ and $C_{\mathrm{VL}}$ for O3 computed using selected MBTA BBH events. The profile log-likelihoods are shown with the addition of $\ln(\mathcal{L}_1)$ (cyan curves) with an arbitrary offset. \label{fig:loglikeli_events_final}}
\end{figure*}

The results for O3 are achieved making a MLE on $\mathcal{L}$ with $C_{\mathrm{HL}} \in[0.875,1.125]$ and $C_{\mathrm{VL}}\in~[0.70,1.30]$. 
They are shown in figure~\ref{fig:loglikeli_events_final}.
From the profile log-likelihoods, we infer the final relative calibration factors as $C_{\mathrm{HL}} =~1.025\pm~0.035$ and $C_{\mathrm{VL}}=~0.916^{+0.096}_{-0.072}$.
They are compatible with $1$ at the $1\sigma$ level, meaning compatible with no relative calibration errors.

As expected from the results performed on simulations, the constraints on $C_{\mathrm{HL}}$ and $C_{\mathrm{VL}}$ using jointly $\mathcal{L}_1$ and $\mathcal{L}_2$ are mostly dominated by the $\mathcal{L}_1$ contribution ($C_{\mathrm{HL}} =~1.025\pm~0.035$, $C_{\mathrm{VL}} = 0.916\pm0.096$). Nevertheless, we get an improvement on the uncertainty of $C_{\mathrm{VL}}$ by including $\mathcal{L}_2$.

Finally, one can also compute the relative calibration factor between Virgo and LIGO Hanford $C_{\mathrm{VH}} =C_{\mathrm{VL}} / C_{\mathrm{HL}}$ and get $C_{\mathrm{VH}} =~0.894^{+0.124}_{-0.101}$.
\section{Conclusion} \label{conclusions}
In this paper, we have investigated a method to measure the relative calibration of the GW strain tested on simulations and applied it to LIGO and Virgo O3 data.
Using GW events detected with MBTA during O3, we measured the relative calibration between the LIGO Hanford and LIGO Livingston detectors with an accuracy of $\sim3.5\%$ ($C_{\mathrm{HL}} =~1.025\pm0.035$) and between the LIGO Livingston and Virgo detectors at the level of $\sim10\%$ ($C_{\mathrm{VL}}=0.916^{+0.096}_{-0.072}$).
It is interesting to note that measurements performed with the Virgo NCal \cite{NCalVirgo:2021} also gave hints of a value of $C_{\mathrm{V}}$ slightly below unity.

The results on the relative calibration factors are promising given the fairly low number of events considered in this study, especially for the detections involving the Virgo detector.
We expect the~uncertainties to be reduced by roughly the square root of the number of events, i.e. by about a factor of~$2$ for O4 given the planned improved sensitivities of the detectors \cite{Prospects:2020}.
This method could also be extended to the cross-calibration of the KAGRA detector.
Another improvement could be to use the reconstructed sky position to constrain the SNR ratio as there will be more events observed by three (or more) detectors.

The relative calibration factors estimated here are averaged over O3.
However, with hundreds of events in a single run as is foreseen for O5 \cite{Prospects:2020} and beyond, or thousands of events with the third-generation detectors \cite{ETscience:2020,CEscience:2017}, this method could be applied on shorter periods to monitor the change in time of the relative calibration between detectors.
Moreover, we could explore the frequency dependence of calibration errors by recomputing the SNR with frequency-dependent scaling factors.
This could be possible for frequency bands that contribute significantly to the SNR of the considered astrophysical signals.
Eventually, it will also be possible to measure calibration errors on timing and phase with better constraints.
\begin{figure*} [h]
\centerline{\includegraphics[trim={0cm 0cm 0cm 0cm},clip,width=1.02\textwidth]{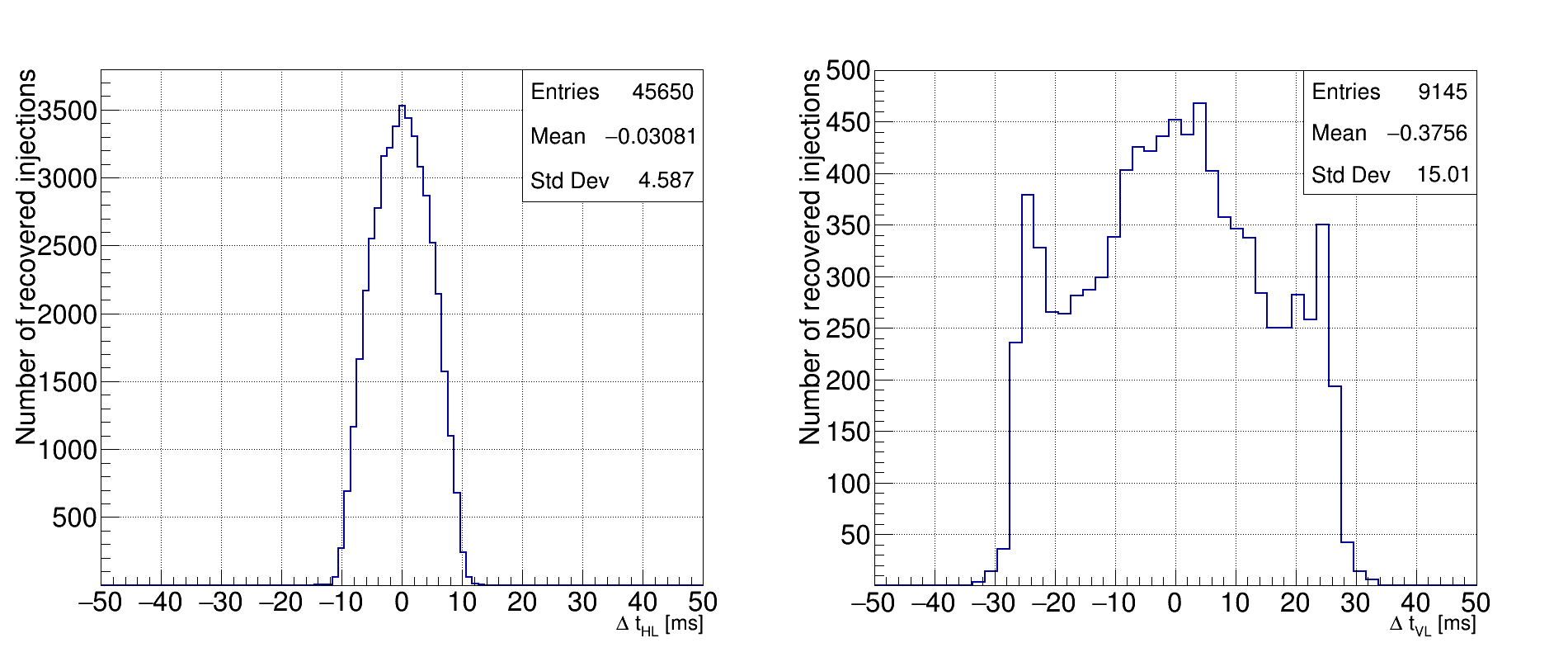}}
\caption{Time of flight distributions for HL ($\Delta t_{\mathrm{HL}}$) and VL ($\Delta t_{\mathrm{VL}}$) recovered injections. \label{fig:distrib_delta_t}}
\end{figure*}
\section*{Acknowledgments}
We thank our LIGO-Virgo collaborators from the CBC and calibration groups for constructive comments. This analysis exploits the resources of the computing facility at the EGO-Virgo site, and of the Computing Center of the Institut National de Physique Nucl\'e{}aire et Physique des Particules (CC-IN2P3/CNRS).
This research has made use of data or software obtained from the Gravitational Wave Open Science Center (gw-openscience.org), a service of LIGO Laboratory, the LIGO Scientific Collaboration, the Virgo Collaboration, and KAGRA.
This material is based upon work supported by NSF's LIGO Laboratory which is a major facility fully funded by the National Science Foundation.
LIGO Laboratory and Advanced LIGO are funded by the United States National Science Foundation (NSF) as well as the Science and Technology Facilities Council (STFC) of the United Kingdom, the Max-Planck-Society (MPS), and the State of Niedersachsen/Germany for support of the construction of Advanced LIGO and construction and operation of the GEO600 detector. Additional support for Advanced LIGO was provided by the Australian Research Council. Virgo is funded, through the European Gravitational Observatory (EGO), by the French Centre National de la Recherche Scientifique (CNRS), the Italian Istituto Nazionale di Fisica Nucleare (INFN) and the Dutch Nikhef, with contributions by institutions from Belgium, Germany, Greece, Hungary, Ireland, Japan, Monaco, Poland, Portugal, Spain.

\newpage
\appendix
\section{Relative phase calibration errors}
We show in figure~\ref{fig:distrib_delta_t} the time of flight distributions of HL and VL recovered injections.
From the standard deviation of these distributions and the number of O3 events we consider in this paper (32 HL and 5 VL), the expected accuracy on the average $\Delta t_{\mathrm{HL}}$ value is $0.8~$ms and on $\Delta t_{\mathrm{VL}}$ is $7~$ms.
These values translate to phase errors of $0.5~$rad and $4.4~$rad at the frequencies of the optimum sensitivity, around $100$~Hz.
These values are much higher than the phase calibration errors given during O3 for LIGO ($<70~$mrad \cite{CalibLIGOO3:2020}) and Virgo ($\sim~35$~mrad with $10~\mu$s on the timing \cite{CalibVirgoO3:2022}).
This shows the limited usefulness of the time of flight for the detectors cross-calibration.
Exploring chirp masses differences from detector-specific parameter estimations could be a way to investigate relative phase calibration errors. \label{appendix}

\end{document}